\documentclass[aps,preprintnumbers,superscriptaddress,letterpaper,twocolumn,nofootinbib]{revtex4}

\usepackage{epsfig,graphics}
\usepackage{amsfonts,amssymb,amsmath} 
\usepackage{amsthm}         

\newcommand{\comment}[1]{}
\newcommand{\ket}[1]{\left | #1 \right\rangle}

\newcommand{\eps}{\varepsilon}
\newcommand{\id}{\openone}
\newcommand{\ot}{\otimes}
\newcommand{\tr}{\text{tr}}
\newcommand*{\cD}{\mathcal{D}}
\newcommand*{\cF}{\mathcal{F}}

\newcommand*{\cS}{\mathcal{S}}
\newcommand*{\cT}{\mathcal{T}}
\def\bibsection{\section*{REFERENCES}} 

\theoremstyle{plain}

\theoremstyle{definition}

\begin{document}
\title{Free randomness can be amplified\footnote{The published version of this work can be found in
    {\it Nature Physics}~{\bf 8}, 450--453 (2012) at \url{http://www.nature.com/nphys/journal/vaop/ncurrent/abs/nphys2300.html}.}}

\author{Roger \surname{Colbeck}}
\email[]{colbeck@phys.ethz.ch}
\affiliation{Institute for Theoretical Physics, ETH Zurich, 8093
Zurich, Switzerland.}
\affiliation{Perimeter Institute for Theoretical Physics, 31 Caroline Street North, Waterloo, Ontario, N2L 2Y5, Canada.}
\author{Renato \surname{Renner}}
\email[]{renner@phys.ethz.ch}
\affiliation{Institute for Theoretical Physics, ETH Zurich, 8093
Zurich, Switzerland.}

\date{\today}

\begin{abstract}
  Are there fundamentally random processes in nature?  Theoretical
  predictions, confirmed experimentally, such as the violation of Bell
  inequalities~\cite{Bell}, point to an affirmative answer.  However,
  these results are based on the assumption that measurement settings
  can be chosen freely at random~\cite{Bell_free}, so assume the
  existence of perfectly free random processes from the outset.  Here
  we consider a scenario in which this assumption is weakened and show
  that partially free random bits can be amplified to make arbitrarily
  free ones.  More precisely, given a source of random bits whose
  correlation with other variables is below a certain threshold, we
  propose a procedure for generating fresh random bits that are
  virtually uncorrelated with all other variables. We also conjecture
  that such procedures exist for any non-trivial threshold. Our result
  is based solely on the no-signalling principle, which is necessary
  for the existence of free randomness.
\end{abstract}
\maketitle

Physical theories enable us to make predictions.  We can ask ``what
would happen if...''\ and reason about the answer, even in scenarios
that would be virtually impossible to set up in
reality~\cite{Bell_free}.  Each scenario corresponds to a choice of
parameters, and it is usually implicitly assumed that any of the
possible choices can be made|the theory prescribes the subsequent
behaviour in every case.  One of the main aims of this Letter is to
identify (minimal) conditions under which such choices can be made
freely, i.e., such that they are uncorrelated with any pre-existing
values (in a precise sense described later).

Free choices are important both at the level of fundamental physics,
and for technological applications.  In almost any cryptographic
protocol, for example, some kind of randomness is needed, and if this
is not generated freely, the protocol can be rendered insecure.  As a
simple example, consider a random number generator used by a casino.
Evidently, a gambler with access to data correlated with these numbers
can exploit this to their advantage.

Another reason why free choices are important is to establish
symmetries on which physical theories can be based.  For example, the
concept of an electron is based on the implicit assumption that we
could pick any of the electrons in the universe and find the same
properties (such as its mass).  More precisely, given a set of
particles that are experimentally indistinguishable, the assumption
that we can sample freely from this set establishes a symmetry between
them.  Following arguments by de Finetti~\cite{deFin,Renner2}, this
symmetry implies that we can treat these particles as independent
particles of the same type.

A scenario in which making free random choices is particularly
relevant is in the context of Bell's theorem. Here, the statistics
produced by freely chosen measurements on an entangled state are used
to conclude that quantum correlations cannot be reproduced by a local
hidden variable theory~\cite{Bell,Bell_free}.  Dropping the assumption
that the measurement settings are freely chosen opens a loophole,
rendering the conclusion invalid.  In particular, if one instead
imagines that the settings were determined by events in the past (this
is sometimes called ``super-determinism'') then it is easy to explain
Bell inequality violations with a local classical model.  However, one
can ask whether the free choice assumption can be relaxed, allowing
for correlations between the measurement settings and other, possibly
hidden, variables, but without allowing their complete
pre-determination.  This has been studied in recent
work~\cite{KPB,Hall_FW,BarrettGisin,Hall_KS,deVit_FW} which shows that
if the choice of measurement settings is not sufficiently free then
particular quantum correlations can be explained with a local
classical model.

This raises the question of whether established concepts in physics
are rendered invalid if one relaxes the (standard) assumption that the
experimenters' choices are perfectly free.  One might imagine, for
example, an experimenter who tries to generate free uniform bits, but
(unbeknown to them) these bits can be correctly guessed with a
probability of success greater than $1/2$ using other (pre-existing)
parameters.  In this Letter, we show that partially free random bits
can be used to produce arbitrarily free ones.  This implies that a
relaxed free choice assumption is sufficient to establish all results
derived under the assumption of virtually perfect free choices.

To arrive at this conclusion we need to make one assumption about the
structure of any underlying physical theory, namely that it is
no-signalling, which essentially implies that local parameters are
sufficient to make any possible predictions within the theory.  As we
explain in the Supplementary Information, it turns out that this
assumption is necessary in order that perfectly free choices can be
consistently incorporated within the theory.\bigskip

In order to describe our result in detail, we need a precise notion of
what partially free randomness is.  The main idea is that, given a
particular causal structure, a variable is \emph{free} if it is
uncorrelated with all other values except those which lie in its
causal future. Our main results are valid independently of the exact
causal structure, but it is natural to consider the causal structure
arising from relativistic space time, which has the property that $Y$
\emph{cannot be caused by} $X$ if $Y$ lies outside the future
lightcone of $X$.

Given a causal structure, we say that $X$ is \emph{perfectly free} if
it is uniformly distributed conditioned on any variable that cannot be
caused by $X$.  This definition, together with the relativistic
understanding of cause above, captures the idea that $X$ is free if
there is no reference frame in which it is correlated with variables
in its past, which corresponds to the notion used by
Bell~\cite{Bell_free}. Note that the definition includes that $X$ is
uniformly distributed, as well as that it is independent of other
values.  While, in other contexts, it may be useful to separate these
properties, in the present work such a distinction is not needed.

We also need a notion of partial freedom.  We say that $X$ is
\emph{$\eps$-free} if it is $\eps$-close in variational distance to
being perfectly free (see Methods).  This measure of closeness is
chosen because of its operational significance: if two distributions
have variational distance at most~$\eps$, then the probability that we
ever notice a difference between them is at most~$\eps$.  As an
example, if a uniformly random bit $X$ is correlated to a pre-existing
bit $W$ such that $P_{X|W=0}(0)=\frac{3}{4}$ and
$P_{X|W=1}(1)=\frac{3}{4}$ then we say that $X$ is $\eps$-free for
$\eps=\frac{1}{4}$.

The idea of the present work is to exploit a particular set of
non-local correlations found in quantum theory that can be quantified
using the chained Bell inequalities~\cite{pearle,BC}.  If we have
perfect free randomness to choose measurements, then the violation of
a Bell inequality indicates that the measurement outcomes cannot be
completely pre-determined~\cite{Bell}.  Bell's arguments have recently
been extended to show that, again under the assumption that we have
perfect free randomness, there is no way to improve on the predictions
quantum theory makes about measurement outcomes~\cite{CR_ext}.  Here,
we show that quantum correlations can be so strong that, even if we
cannot choose the measurements perfectly freely, the outputs are
nevertheless perfectly free.

\begin{figure}
\includegraphics[width=0.5\textwidth]{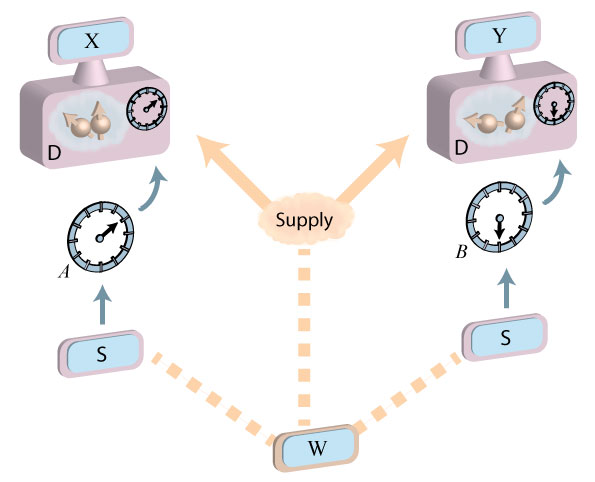}
\caption{{\bf Illustration of the bipartite setup.}  Spacelike
  separated measurements are carried out using devices denoted D.  The
  choices of measurement, $A$ and $B$, are derived from bits generated
  by two sources of weak randomness, denoted S. These bits are only
  partially free, i.e., they may be correlated (represented by the
  dashed line) with each other and with some other variables $W$ (to
  be interpreted as parameters provided by a possible higher theory),
  which may also influence the supply of states being measured.  By
  exploiting correlations between the outcomes, $X$ and $Y$, we show
  that, in spite of the lack of perfectly free randomness to choose
  settings, the outcome $X$ is arbitrarily close to being uniform and
  uncorrelated with $W$.}
\label{fig:setup}
\end{figure}

To generate these correlations, we consider an experimental setup
where local measurements are performed on a pair of maximally
entangled qubits (see Fig.~\ref{fig:setup}). We first make the
(temporary) assumption that the joint distribution of measurement
outcomes conditioned on the choices, $P_{XY|AB}$, is the one predicted
by quantum theory for this setup.  Crucially, however we do not
require completeness of quantum theory, i.e., that quantum theory is
maximally informative about the measurement outcomes.  Instead, we
consider arbitrary additional parameters, $W$, that may be provided by
a higher theory. Within this setup, our assumptions can
be stated as follows.\smallskip\\
{\bf NS:} $P_{XY|AB\,W=w}$ is no-signalling for all $w$ (i.e., $P_{X|AB\,W=w} = P_{X|A\,W=w}$ and $P_{Y|AB\,W=w} = P_{Y|B\,W=w}$).\smallskip\\
{\bf QT:} $P_{XY|AB}$ is that predicted by quantum theory.\smallskip

Our first main result is that, under the above assumptions, there
exists a protocol that uses sources of $\eps$-free bits to generate
arbitrarily free bits for any $\eps<(\sqrt{2}-1)^2/2\approx 0.086$
(see Theorem~1 in the Methods).

It is natural to ask whether the assumption that quantum theory
correctly predicts the correlations (assumption~QT), is necessary, or
whether, instead, the presence of sufficiently strong correlations can
be \emph{certified} using $\eps$-free bits.  By certification, we mean
a procedure to test the correlations such that it is essentially
impossible that the test passes without the generated bits being
arbitrarily free.  This is also relevant in a cryptographic context,
where the states and measurements are not trusted, and could have been
chosen by an adversary with partial knowledge, $W$, of the measurement
settings.

Our second main result is that, under assumption~NS alone, there
exists a protocol that uses $\eps$-free bits to certify the generation
of arbitrarily free bits for any $\eps<0.058$ (see Theorem~2 in the
Methods).  In other words, there exists a device-independent protocol
for free randomness amplification.  Clearly this second scenario,
where the assumption that the correlations are those predicted by
quantum theory is dropped, is more demanding, hence the smaller range
of $\eps$ for which free randomness amplification is successful.
Nevertheless, the fact that it is possible at all is already
fascinating.

It is an open question as to how far the threshold on $\eps$ can be
pushed such that free randomness amplification remains possible (in
either scenario).  It turns out that using chained Bell correlations
there is a limit, since (as shown in the Supplementary Information)
for $\eps\geq(1-1/\sqrt{2})/2\approx 0.146$, these
correlations admit a local classical explanation.  However, we
conjecture that there exist protocols based on other correlations such
that for any $\eps<\frac{1}{2}$, $\eps$-free bits can be used to
generate arbitrarily free bits.  We give some evidence for this in the
Supplementary Information.

\bigskip

Before discussing the implications of these results, we first remark
that the use of no-signalling conditions for information processing
tasks was first observed in~\cite{BHK} in the context of key
distribution.  We also note that what we call free randomness has
sometimes been called ``free will'' in the literature
(e.g.~\cite{ConKoc_FW1,ConKoc_FW2}).  In this language, we could
restate our main result as a proof that free will can be amplified.

A sequence of bits $S_1, S_2 \ldots$ for which each $S_i$ is
$\eps$-free is known in classical computer science as a
Santha-Vazirani source~\cite{SanthaVazirani}.  It has been shown that
no classical algorithm can extract even a single uniform bit from such
a source (without an additional seed; we elaborate on this point in
the Supplementary Information). In contrast, our main result implies
that such a bit can be generated using a quantum algorithm.

It is worth comparing randomness amplification, as considered here,
with \emph{randomness expansion}, introduced in~\cite{ColbeckThesis}
and further developed in~\cite{PAMBMMOHLMM,CK2}.  There, an initial
perfectly random finite seed is used within a protocol to generate a
longer sequence of random bits using untrusted devices.  By contrast,
we do not require such a seed in the present work, but instead have an
arbitrarily large supply of imperfect randomness.

A potential application of our protocol is as a method for generating
a seed, to be used with an extractor to extract further randomness
from a partially free source, or to seed a randomness expansion
protocol.  Using Trevisan's extractor~\cite{Trevisan,DPVR}, for
example, in the first case we could generate random bits at the
entropy rate of the partially free source.  In the second case,
provided that the protocols can be securely composed, a secure
randomness expansion protocol may allow a large amount of free
randomness to be derived from a finite number of uses of partially
free sources.

We also comment on the implications of our result for experimental
demonstrations of Bell-inequality violations.  There are several
potential loopholes in current experiments, leaving the door open for
die-hards to reject certain philosophical implications.  One such
loophole that has received only minor attention in the literature is
the so called \emph{free-choice loophole}, which has been addressed in
a recent experiment~\cite{Scheidl&}.  This loophole says that the
supposedly free measurement settings were in fact correlated with the
entanglement source (perhaps via some hidden system).  In the
aforementioned experiment, this is addressed by using random number
generators, triggered at spacelike separation from the source of
entangled pairs.  However, as acknowledged in~\cite{Scheidl&}, this
leaves room for ``super-determinism'', since it is impossible to
exclude the possibility that the random number generator and the
source of entanglement are correlated via an additional hidden system.

Use of our result is also not able to close this loophole, and, since
we can never rule out that the universe is deterministic, we don't see
any way to completely close it.  Nevertheless, our result complements
existing work on the weakening of free choice in Bell
experiments~\cite{KPB,Hall_FW,BarrettGisin,Hall_KS,deVit_FW}: instead
of having to assume that the entanglement source and the random number
generator are completely uncorrelated, we would only need to assume
that they are not strongly correlated.  Furthermore, if our conjecture
is true (i.e., $\eps$-free bits can be amplified for any $\eps < 1/2$)
then, for certain Bell tests, it would be sufficient to assume only
that the source and the random number generator are not
\emph{completely} correlated.

In other scenarios in which the assumption of free choice is critical,
generating such choices via our free randomness amplification
procedure would also enable stronger conclusions to be drawn.  For
example, within classical cryptography a wide range of cryptographic
tasks that use perfect randomness are rendered impossible if the
parties performing them have access only to imperfect randomness
sources~\cite{Dodis&}.  Our result shows that in a quantum setting,
this is not the case; any task that can be performed securely using
perfect randomness can also be performed securely with access only to
(sufficiently free) imperfect randomness.

\section*{Methods}

In this section we give more technical versions of our definitions and
main results.\smallskip

We consider a set $\Gamma$ that includes all random variables of
interest in our setup (see Fig.~\ref{fig:setup}) and equip $\Gamma$
with a causal structure (mathematically, this is a preorder relation
between its elements). As explained in the main text, it is convenient
(but not necessary) to think of the causal structure induced by
relativistic space time. \smallskip

\noindent\emph{Definition}|Let $X \in \Gamma$ and 
let $\Gamma_X$ be the subset of random variables from $\Gamma$ that
cannot be caused by $X$ (in particular, $\Gamma_X$ does not include
$X$). Then $X$ is called \emph{$\eps$-free} if
\begin{equation}\label{eq:epsfree}
  D(P_{X|\Gamma_X=\gamma_X},P_{\bar{X}})\leq\eps,
\end{equation}
for all $\gamma_X$, where $P_{\bar{X}}$ denotes the uniform
distribution on $X$.  $D(\cdot)$ denotes the
\emph{variational distance}, defined by
$D(P_X,Q_X):=\frac{1}{2}\sum_x|P_X(x)-Q_X(x)|$. (Here and in the following we use lower case to denote particular
instances of upper case random variables.) \smallskip

\begin{figure}
\includegraphics[width=0.5\textwidth]{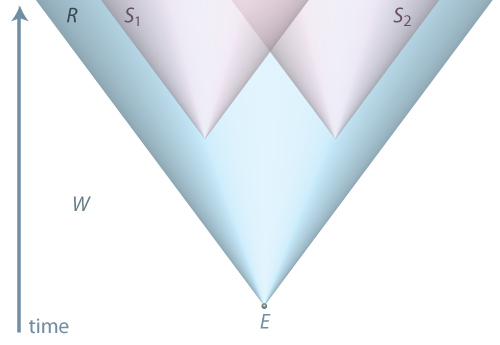}
\caption{{\bf Typical causal structure of a protocol.} A randomness
  amplification protocol for generating a random bit $R$ may be
  initiated at a particular location and time represented by a spacetime
  point $E$. Depending on the protocol, information correlated to $R$
  may be generated at various locations within the causal future of
  $E$, depicted by the blue region. Any point in this region may
  therefore potentially be in the causal future of $R$.  The bit $R$
  satisfies our definition of being \emph{free} if it is uncorrelated
  with anything outside this region, indicated by the variable
  $W$. The protocol may invoke sources of random bits $S_1$ and $S_2$
  at many locations. Technically, the only requirement on the causal
  structure is that the causal futures of each of these source bits
  (pink regions) lie in the blue region.}
\label{fig:spacetime}
\end{figure}

For our main claims, we use random variables $S_i \in \Gamma$ for
$i=1,2,\ldots$ (these denote the random bits generated by an
$\eps$-free source) and $R \in \Gamma$ (the random bit generated by
the protocol), where the causal structure can be arbitrary up to the
following constraint: the causal future of $R$ includes the causal
future of any $S_i$ (see Fig.~\ref{fig:spacetime}). \smallskip

\noindent\emph{Theorem~1}|There exists a protocol that takes as input
$S_i$ and outputs $R$ such that the following holds
under the assumptions~NS and~QT: if $S_i$ are $\eps$-free, for any
$\eps<(\sqrt{2}-1)^2/2 \approx0.086$, then $R$ is arbitrarily free,
except with arbitrarily small probability.\smallskip

The proof relies on the bipartite setup of Fig.~\ref{fig:setup}.  It
is parameterized by an integer, $N$, corresponding to the number of
measurement settings on each side.  $A\in\{0,2,\ldots,2N-2\}$ and
$B\in\{1,3,\ldots,{2N-1}\}$ correspond to the choices of measurements
and $X\in\{+1,-1\}$ and $Y\in\{+1,-1\}$ are their respective outcomes.
We introduce a measure of the strength of the resulting correlations
by defining
\begin{eqnarray}
I_N:=P(X=Y|a_0,b_0)+
\!\!\!\sum_{\genfrac{}{}{0pt}{}{a,b}{|a-b|=1}}
P(X\neq Y|a,b)\, ,\label{eq:IN}
\end{eqnarray}
where $a_0=0$, $b_0=2N-1$ and $P(X\neq Y|a,b)$ is the probability that
the measurements give different outcomes for settings $A=a$, $B=b$.
This quantity was originally introduced to study \emph{chained Bell
  correlations}~\cite{pearle,BC}, and has found use in
cryptography~\cite{BHK,BKP} and quantum
foundations~\cite{ColbeckRenner,CR_ext}.  It turns out that all
classical correlations (i.e., those that can be reproduced from
classical shared randomness) satisfy $I_N\geq 1$, while quantum
correlations exist for which
\begin{equation}\label{eq:INQM}
I_N=2N\sin^2\frac{\pi}{4N},
\end{equation}
which tends to~$0$ in the limit of large $N$ (the state and measurements
required to achieve this are given in the Supplementary Information).

In the proof of Theorem~1, we use the following lemma about
no-signalling distributions.  This lemma bounds the independence of
the output bits from the choices $A$ and $B$ as well as any values
$W$. The bound is given in terms of the strength of quantum
correlations, quantified using $I_N$, and how free the measurement
settings are, quantified via
$$q_N(a,b):=\min_{\genfrac{}{}{0pt}{}{a',b',w'}{|a'-b'|=1}}\left[\frac{P_{W|a'b'}(w')}{P_{W|ab}(w')}\right].$$
\smallskip

\noindent\emph{Lemma 1}|If $P_{XY|ABw}$ is no-signalling for all
$w$, and $q_N(a,b)>0$, then
\begin{eqnarray}\label{eq:mainX}
  D(P_{XW|ab},P_{\bar{X}}\times P_{W|ab})\leq \frac{I_N}{2q_N(a,b)}
\end{eqnarray}
for all $a$ and $b$, where $P_{\bar{X}}$ denotes the uniform
distribution on $X$.\smallskip

The proof of this lemma is given in the Supplementary Information.\smallskip

\noindent\emph{Proof of Theorem~1}|The protocol relies on the
correlations introduced above, where the source bits $S_i$ are used to
choose $A$ and $B$, and where $X$ is taken as the final output $R$.
It remains to show that $R$ is arbitrarily free in the limit of large
$N$.  Let $W$ be any subset of $\Gamma$ that is not in the causal
future of $R$, and therefore, by assumption, not in the causal future
of the source bits $S_i$.  Note that
$$q_N(a,b)=\min_{\genfrac{}{}{0pt}{}{a',b',w'}{|a'-b'|=1}}\left[\frac{P_{AB|w'}(a',b')}{P_{AB|w'}(a,b)}\right]$$
in the case of uniform $P_{AB}$, which we can assume without loss of
generality.  For $N=2^r$, the measurement settings, $A$ and $B$ can be
picked using $r$ $\eps$-free source bits, and hence
$q_{2^r}(a,b)\geq\left(\frac{1-2\eps}{1+2\eps}\right)^{2r}.$ Inserting
this into~\eqref{eq:mainX} gives
$$D(P_{XW|ab},P_{\bar{X}}\times P_{W|ab})\leq \frac{I_{2^r}}{2}\left(\frac{1+2\eps}{1-2\eps}\right)^{2r}.$$

Substituting the value of $I_{2^r}$ obtainable in quantum
theory (see~Eq.~\eqref{eq:INQM}) gives
$$D(P_{XW|ab},P_{\bar{X}}\times P_{W|ab})\leq2^r\left(\frac{1+2\eps}{1-2\eps}\right)^{2r}\sin^2\left(\frac{\pi}{2^{r+2}}\right).$$
Hence, using the bound $\sin x\leq x$ for $x\geq 0$, it follows that
$$D(P_{XW|ab},P_{\bar{X}}\times
P_{W|ab})\leq\frac{\pi^2}{16}\left(\frac{1+2\eps}{\sqrt{2}(1-2\eps)}\right)^{2r}=:\delta_{r,\eps}\,
,$$
which tends to $0$ as $r$ tends to infinity provided
$\eps<(\sqrt{2}-1)^2/2$.

Note that $D(P_{XW|ab},P_{\bar{X}}\times P_{W|ab})$ is equal to
$\sum_wP_{W|ab}(w)D(P_{X|abw},P_{\bar{X}})$, i.e., the expectation
over $W$ of the amount by which the output bits are free.  Using
Markov's inequality, we have that $D(P_{X|abw},P_{\bar{X}})<\alpha$,
except with probability at most $\delta_{r,\eps}/\alpha$, for any
$\alpha>0$.  Thus, taking $\alpha=\sqrt{\delta_{r,\eps}}$, if the
initial sources are $\eps$-free for $\eps<(\sqrt{2}-1)^2/2$, then, in
the limit of large $r$, their outputs are
$\sqrt{\delta_{r,\eps}}$-free, except with probability
$\sqrt{\delta_{r,\eps}}$.  The claim then follows because
$\delta_{r,\eps}$ can be made arbitrarily small by choosing a
sufficiently large $r$. \qed\bigskip

In the second part of our main result, we show that assumption~QT can
be omitted. \smallskip

\noindent\emph{Theorem~2}|There exists a protocol that takes as input
$S_i$ and outputs $R$ such that the following holds
under the assumption~NS: if $S_i$ are $\eps$-free, for any
$\eps<0.058$, then $R$ is certified to be
arbitrarily free, except with arbitrarily small
probability.\smallskip

We give a specific protocol that achieves this task 
and analyse it in the Supplementary Information.

For completeness, we state our conjecture:\smallskip

\noindent\emph{Conjecture~1}|The restriction on $\eps$ in Theorems~1
and~2 can be replaced by $\eps<\frac{1}{2}$.\smallskip

It is likely that these alternative protocols need to go beyond the
bipartite setup to succeed, as discussed in the Supplementary
Information.


\section*{Acknowledgements}
We thank Viktor Galliard for useful discussions and L\'idia del Rio
for the figures. Research at Perimeter Institute is supported by the
Government of Canada through Industry Canada and by the Province of
Ontario through the Ministry of Research and Innovation.  R.R.\
acknowledges support from the Swiss National Science Foundation (grant
No.\ 200020-135048, the NCCR QSIT, and the CHIST-ERA project DIQIP)
and from the European Research Council (grant No.\ 258932).

\begin{widetext}
\appendix
\def\bibsection{\section*{SUPPLEMENTARY REFERENCES}} 

\makeatletter
\def\tagform@#1{\maketag@@@{(S\ignorespaces#1\unskip\@@italiccorr)}}
\makeatother

\vspace{-1cm}
\setcounter{equation}{0}
\section*{SUPPLEMENTARY INFORMATION}

\subsection*{Correctness vs. completeness of quantum theory}
For a large part of this work, we make the assumption that quantum
theory is correct, but not necessarily complete.  To recap,
correctness means that the observed distributions of measurement
outcomes will follow those given by quantum theory. Completeness is a
stronger notion and means that there is no higher theory which better
explains the outcomes (i.e., a theory that provides us with additional
information allowing us to better predict the outcomes of
measurements).  While the correctness assumption is often sufficient,
for instance, to predict the behaviour of a physical device, there are
scenarios in which the additional assumption of completeness is
crucial.  In quantum cryptography, for example, the aim is to show
that there is no attack within the laws of physics that renders a
cryptographic scheme insecure.

The distinction between these two notions is a significant one.
Correctness is an operational concept and is in principle
experimentally verifiable: by repeatedly measuring a system, we can
place increasingly accurate bounds on its statistics.  Completeness,
on the other hand, is not directly verifiable, and many of our
experimentally well-founded theories have found higher explanation in
the past (statistical mechanics explains many phenomena in
thermodynamics at a higher level, for example).  We therefore argue
that assuming completeness is a different class of assumption, and we
do not make it in this work.

Note that in the second part of our result, we also drop the
assumption of correctness, showing that by performing measurements
(chosen with $\eps$-free bits) we can verify the correlations to a
sufficient level to conclude that their outcomes are random.

\subsection*{General protocols and adversarial scenario}
Here we explain the setup of a general protocol for free randomness
amplification.  Such a protocol involves performing $M$ separated
measurements for some $M\geq 2$ (by separated we mean that the
no-signalling conditions hold as defined below).  Each measurement
choice, $A_i$, is derived from a number of $\eps$-free bits $S_j$, and
$X_i$ is the corresponding outcome.  The idea is that, for an
appropriate distribution $P_{X_1\ldots X_M|A_1\ldots A_M}$ specified
by the protocol (and realizable using a set of measurements on a
quantum system prepared in a certain state), one (or more) of the
output bits $X_i$, or some function of them, is arbitrarily close to
being perfectly free.  As in the main text, we require that this holds
within any causal structure such that the potential causal future of
the protocol's output, $R$, includes the causal future of all the
$S_j$.

We can also recast the setup in an adversarial scenario.  Here, the
set of variables, $W$, with which the partially free sources may be
correlated, can be thought of as being held by an adversary. (More
generally, one could think of a correlated system (instead of the set
$W$) which takes an input, corresponding to a choice of measurement,
and gives an output (analogous to a quantum system).  However, both
the input and output can be included in the set $W$.)  The adversary
then supplies an $M$-party system whose behaviour, $P_{X_i\ldots
  X_M|A_i\ldots A_Mw}$, may depend on $W$.  For the $M$-party setup,
we need to slightly generalize the assumptions given in the main text
for the bipartite case:
\begin{itemize}
\item[$\text{NS}_M$:] $P_{X_1\ldots X_M|A_1\ldots A_Mw}$ is no-signalling for
all $w$. (An $M$-party distribution $P_{X_1\ldots X_M|A_1\ldots A_M}$
is no-signalling if $P_{X^\bot_i|A_1\ldots A_M}=P_{X^\bot_i|A^\bot_i}$
for all $i$, where
$X^\bot_i:=X_1\ldots X_{i-1}X_{i+1}\ldots X_M$.)
\item[$\text{QT}_M$:] There exists an $M$-party quantum state, $\rho$,
  and POVMs $\{E^{a_1}_{x_1}\}$, $\{F^{a_2}_{x_2}\}$, $\ldots$
  (positive operators that satisfy $\sum_{x_i}E^{a_i}_{x_i}=\id$ for
  all $a_i$ etc.) such that $P_{X_1\ldots X_M|a_1\ldots a_M}(x_1,\ldots,
  x_m)=\tr\left((E^{a_1}_{x_1}\ot F^{a_2}_{x_2}\ot\ldots)\rho\right)\,
  .$
\end{itemize}

For the first part of our result (Theorem~1), we consider the case
where the adversary sets up this system such that, if the partially
free bits are used to choose $\{A_i\}$, the resulting distribution
(averaged over $W$) is indistinguishable from that generated by
performing quantum measurements on a quantum state specified by the
amplification protocol, i.e., we assume $\text{QT}_M$ holds.  The
adversary is further restricted by assumption~$\text{NS}_M$.  The
output of the protocol is then considered free if it is uniformly
distributed and the adversary is unable to learn anything about it.

In the second part of the result (Theorem~2), the adversary is no
longer required to ensure that the distribution appears quantum, but
is restricted only by assumption~$\text{NS}_M$.  In this case, certain
statistical tests (prescribed by the protocol) will be performed on
the outcomes of measurements on the supplied states.  These tests
should have the property that no state can pass them without the
generated bits being arbitrarily free.

\subsection*{Necessity of the no-signalling conditions}
Here we show that, in order to incorporate perfectly free random bits
into a theory, 
it is necessary that this theory satisfies the no-signalling
conditions.  The argument essentially follows one proposed
in~\cite{CR_ext} (see also~\cite{CRcomment}).\smallskip

\noindent\emph{Lemma}|Let $W$, $A_i$ and $X_i$ be random variables for
$i\in \{1, \ldots, M\}$ such that $W$, $A_j$ and $X_j$ do not lie in
the causal future of $A_i$ for all $i$ and $j\neq i$.  If, for all
$i$, $A_i$ is perfectly free, then $P_{X_1\ldots X_M|A_1\ldots A_Mw}$
is no-signalling.\smallskip

\noindent\emph{Proof}|Using Bayes' rule we have
$$P_{X_i^\bot|A_1\ldots A_Mw}=P_{X_i^\bot|A_i^\bot
  w}\frac{P_{A_i|A_i^\bot X_i^\bot w}}{P_{A_i|A_i^\bot w}}\, .$$
Since, by assumption, $A_i$ is perfectly free, and $W$, $A_i^\bot$ and
$X_i^\bot$ are not in the causal future of $A_i$, we have
$P_{A_i|A_i^\bot X_i^\bot w}= P_{A_i|A_i^\bot w}=P_{A_i|w}$.  It hence
follows that $P_{X_i^\bot|A_1\ldots A_Mw}=P_{X_i^\bot|A_i^\bot w}$,
i.e.\ that $P_{X_1\ldots X_M|A_1\ldots A_Mw}$ is no-signalling.
\qed


\subsection*{Implications for randomness extraction}

A further application of our result is in the context of randomness
extraction, introduced in~\cite{BBR,ILL}.  This is the task of taking
a string of bits about which there may be some side information and
using it to generate a string which is uniform even given this side
information.  All previous protocols for this task were classical and
require an additional uniform random seed (i.e.\ bits which are
perfectly free) which acts as a catalyst (there have recently been
extensions of this work to certain imperfect seeds, provided they are
uncorrelated with the string being compressed and that the size of the
side information is bounded~\cite{KK}).  In the case without an
independent seed, it has been shown that no classical algorithm can
extract even a single uniform bit from an adversarially controlled
string of partially free bits~\cite{SanthaVazirani,Dodis&}.  This
shows that free randomness amplification, which we show is possible in
this work, cannot be done using only classical (deterministic)
information processing.


\subsection*{Use of multiple independent sources}
One may wonder whether, given the no-signalling assumption, the
outputs of separated partially free sources are necessarily
independent of each other.  If this were the case, i.e., if the
sources were independent, it would be possible to generate arbitrarily
free bits by a purely classical procedure.  More precisely, as shown
in~\cite{Vazirani}, two \emph{independent} sources of $\eps$-free bits
can be used to generate $\eps'$-free ones for any $\eps'>0$ by
generating sufficiently long strings of instances from each source and
outputting the number of places in which both strings take value 1
modulo 2 (the GF(2) inner-product of the output strings).

In the following, we argue that the no-signalling assumption does not
generally imply that separated sources produce independent outputs.
This is easily seen in the following example in which the bits are
$\eps$-free for a causal structure where the second bit is in the
future of the first.  Suppose that the separated sources share the
quantum state
$$(\frac{1}{2}+\eps)\ket{\uparrow\uparrow}+\sqrt{\frac{1}{4}-\eps^2}\ket{\uparrow\downarrow}+(\frac{1}{2}-\eps)\ket{\downarrow\uparrow}+\sqrt{\frac{1}{4}-\eps^2}\ket{\downarrow\downarrow}$$
and generate their partially free bits by measurement in the
$\{\ket{\uparrow},\ket{\downarrow}\}$ basis.  The resulting
distribution corresponds in effect to choosing the bias of the second
bit depending on the output of the first. However, by construction,
this source of randomness is clearly no-signalling (because quantum
theory has this property).

While the classical construction of~\cite{Vazirani} is not in general
applicable to such correlated sources, our result shows that the
partially free output of the sources can nevertheless be turned into
almost perfectly free uniform randomness.



\subsection*{Proof of Lemma 1}
We remark that this proof is a generalization of one given
in~\cite{CR_ext}, which in turn was based on a series of
work~\cite{ColbeckRenner,BKP} going back to the first provably secure
device-independent key distribution protocol~\cite{BHK}.

Recall that we are working in a bipartite setup where separate
measurements are made with choices $A\in\{0,2,\ldots,2N-2\}$ and
$B\in\{1,3,\ldots,{2N-1}\}$, and outcomes $X\in\{+1,-1\}$ and
$Y\in\{+1,-1\}$. We first consider the quantity $I_N$ (defined in the
main text) evaluated for the conditional distribution
$P_{XY|ABw}=P_{XY|AB\,W=w}$,
for any fixed $w$. The idea is to use this quantity to bound the trace
distance between the conditional distribution $P_{X|aw}$ and its
negation, $1-P_{X|aw}$, which corresponds to the distribution of $X$
if its values are interchanged.  If this distance is small, it follows
that the distribution $P_{X|aw}$ is roughly uniform.

For $a_0 = 0$, $b_0= 2N-1$, we have
\begin{align} 
I_N(P_{XY|ABw})
&:=P(X=Y|a_0,b_0,w)+\!\!\!\sum_{\genfrac{}{}{0pt}{}{a,b}{|a-b|=1}}P(X\neq Y|a,b,w) \nonumber   \\
&\geq D(1-P_{X|a_0 b_0 w},P_{Y|a_0 b_0 w}) +\!\!\!
\sum_{\genfrac{}{}{0pt}{}{a,b}{|a-b|=1}}D(P_{X|abw},P_{Y|abw})
\nonumber\\
&=D(1-P_{X|a_0w},P_{Y|b_0w}) +\!\!\!
\sum_{\genfrac{}{}{0pt}{}{a,b}{|a-b|=1}}D(P_{X|aw},P_{Y|bw})
\nonumber\\
&\geq D(1-P_{X|a_0w},P_{X|a_0w})\nonumber \\
&=2D(P_{X|a_0b_0w}, P_{\bar{X}}) \, , \label{eq:6}
\end{align}
where we have used the relation $D(P_{X|\Omega}, P_{Y|\Omega}) \leq
P({X \neq Y | \Omega})$ for any event $\Omega$, the no-signalling
conditions $P_{X|abw}=P_{X|aw}$ and $P_{Y|abw}=P_{Y|bw}$ and the
triangle inequality for $D$.

Since the quantity $I_N(P_{XY|ABw})$ cannot be computed without
access to $w$, we instead consider
\begin{eqnarray}
  I_N(P_{XY|AB})&:=&P(X=Y|a_0,b_0)+\!\!\!\sum_{\genfrac{}{}{0pt}{}{a,b}{|a-b|=1}}
  P(X\neq Y|a,b) \nonumber
\\
  &=&\sum_{w}P_{W|a_0b_0}(w)P(X=Y|a_0b_0w)+\!\!\!\sum_{\genfrac{}{}{0pt}{}{a,b}{|a-b|=1}}\sum_{w}\frac{P_{W|ab}(w)}{P_{W|a_0b_0}(w)}P_{W|a_0b_0}(w)P(X\neq
  Y|abw)\nonumber
\\
  &\geq&\sum_{w}P_{W|a_0b_0}(w)\min_{\genfrac{}{}{0pt}{}{a,b,w}{|a-b|=1}}\left[\frac{P_{W|ab}(w)}{P_{W|a_0b_0}(w)}\right]I_N(P_{XY|ABw})\nonumber\\
  &\geq&2\min_{\genfrac{}{}{0pt}{}{a,b,w}{|a-b|=1}}\left[\frac{P_{W|ab}(w)}{P_{W|a_0b_0}(w)}\right]
  D(P_{XW|a_0b_0},P_{\bar{X}}\times P_{W|a_0b_0})\, .
\end{eqnarray}

Using the definition of $q_N(a,b)$ (see the main text), we have the
bound
$$D(P_{XW|a_0b_0},P_{\bar{X}}\times
P_{W|a_0b_0})\leq\frac{I_N(P_{XY|AB})}{2q_N(a_0,b_0)}\, .$$
The proof for arbitrary $a$ and $b$ (rather than $a=a_0$, $b=b_0$)
follows by symmetry. 

\section*{Proof of Theorem~2}

To prove Theorem~2, we consider an explicit randomness amplification
protocol, $\Pi_N$, as depicted below. The protocol is basically a
one-party version of a secret key distribution protocol proposed by
Barrett, Hardy and Kent~\cite{BHK}. The protocol $\Pi_N$ depends on a
parameter, $N \in \mathbb{N}$, that determines the quality of the
amplified randomness it generates. During its run, $\Pi_N$ accesses
two separated measurement devices, $\cD_A$ and $\cD_B$, as well as two
sources of $\eps$-free bits, $\cS_A$ and $\cS_B$, located next to the
devices $\cD_A$ and $\cD_B$, respectively. A priori, nothing is
assumed about these devices except that $\cD_A$ (or $\cD_B$), on input
$\alpha$ (or $\beta$) from the interval $[0, 2 \pi]$, produces an
output $X$ (or $Y$) from the set $\{-1, 1\}$.  Ideally, however, the
devices $\cD_A$ and $\cD_B$ reproduce the statistics obtained from
Bell measurements, i.e., local measurements on a bipartite system in
state $(1/\sqrt{2})(\ket{\uparrow} \otimes
\ket{\uparrow}+\ket{\downarrow} \otimes \ket{\downarrow})$ with
respect to measurement bases $\{\ket{\alpha}, \ket{\alpha + \pi} \}$
and $\{\ket{\beta}, \ket{\beta + \pi} \}$, respectively, where
$\ket{\theta} :=
\cos\frac{\theta}{2}\ket{\uparrow}+\sin\frac{\theta}{2}\ket{\downarrow}$,
so that
\begin{align} \label{eq_QMstatistics}
  P(X \neq Y) = \sin^2\frac{\alpha - \beta}{2} \, .
\end{align}
The protocol $\Pi_N$ perform tests to check whether the statistics
obtained from the devices $\cD_A$ and $\cD_B$
satisfies~\eqref{eq_QMstatistics}. Only if these tests succeeds, the
protocol outputs a random bit $R \in \{0,1\}$, which is then
guaranteed to be almost perfectly free (as shown below). Otherwise,
the protocol simply aborts (in which case we set $R = \perp$).  For
convenience, we define $M=N^{299/100}$ and take $N$ to be such that
$\log N$ and $\log(M/N)$ are integers (e.g., by taking
  $N=2^{100r}$ for some integer $r$).

\begin{center}
 \framebox[18cm][l]{\begin{minipage}{17cm}
  \smallskip
  {\bf \hspace{0.1cm} Protocol $\Pi_N$}
  \begin{enumerate}
  \item Repeat the following for all $q = 1, \ldots, M$: Invoke
    $\cD_A$ and $\cD_B$ with inputs $\alpha = \frac{\pi}{2N} A_q$ and
    $\beta = \frac{\pi}{2N} B_q$, respectively, where $A_q \in \{0, 2,
    \ldots, 2N -2\}$ and $B_q \in \{1, 3, \ldots, 2 N-1\}$ are chosen
    at random, using bits from the sources $\cS_A$ and $\cS_B$,
    respectively, and record the outcomes $X_q \in \{-1, 1\}$ and $Y_q
    \in \{-1, 1\}$, respectively.
  \item \label{step:modS} Define the set $\cT := \{q : \, |A_q - B_q| = 1
    \, \text{ or } \, (A_q, B_q) = (0, 2N-1) \}$. Check that the
    cardinality $|\cT|$ of $\cT$ satisfies $|\cT| \in [M/N, 3
    M/N]$. If this test fails then set $R = \perp$ and abort.
  \item \label{step:check} For each $q \in \cT$ check that $X_q = Y_q$
    (if $(A_q, B_q) \neq (0, 2N-1)$) or $X_q \neq Y_q$ (otherwise). If
    for at least one $q \in \cT$ this test fails then set $R = \perp$
    and abort.
  \item \label{step:output} Choose $f \in \cT$ at random, using bits
    from the source $\cS_A$, and output $R=X_f$.
  \end{enumerate}
  \smallskip
\end{minipage}}
\end{center}

We remark that our aim is to provide a proof of principle for the
possibility of randomness amplification. The specific protocol $\Pi_N$
we use, as well as its analysis, are therefore not optimized in terms
of the dependence on parameters such as the quality $\eps$ of the
initial randomness or the number of times the measurement devices need
to be used. Rather, the parameters are chosen such that it is
convenient to verify the claim below.

\bigskip

\noindent {\bf Claim.}
If the random bits used by the protocol $\Pi_N$ are $\eps$-free, for
$\eps < 0.058$, then the following statements hold:
  \begin{itemize}
  \item For any $\eps'>0$, any behaviour of $\cD_A$ and $\cD_B$, and
    under the non-signalling assumption~NS (see the main text), the
    probability (over the outputs of the $\eps$-free sources) that the
    protocol $\Pi_N$ does not abort and outputs a bit $R\in\{0,1\}$
    that is not $\eps'$-free tends to $0$ as $N$ tends to infinity,
    i.e.,
    \begin{align*}
      \lim_{N \to \infty} P(R\neq\perp \, \wedge \, R
      \text{ not $\eps'$-free}) = 0 \, .
     \end{align*}
   \item If the statistics of $\cD_A$ and $\cD_B$
     satisfy~\eqref{eq_QMstatistics} then the probability (over the
     outputs of the $\eps$-free sources) that the protocol $\Pi_N$
     aborts tends to $0$ as $N$ tends to infinity, i.e.,
  \begin{align*}
    \lim_{N \to \infty} P(R=\perp) = 0 \, .
  \end{align*}
  \end{itemize} 

\begin{proof}
  Throughout the proof we will use that, whenever a value (e.g., $A
  \in \{0, 2, \ldots, 2N-2\}$) is chosen at random (using $\eps$-free
  bits) from a set of size $N$, then the probability $P_{A|w}(a)$ of
  any possible value $a$ conditioned on any value of the additional
  information $w$ (as well as all other pre-existing values) is
  contained in the interval $[({1/2 - \eps})^{\log N}, ({1/2 +
    \eps})^{\log N}]$. In the remainder of this proof, we consider a
  fixed $w$ and all probabilities are conditioned on $w$ (which we
  sometimes omit to simplify the notation).

  To prove the first part of the claim, we set $I_N^* = 2 N^{-1/100}$
  and define
  \begin{align*}
    \cF := \{q \in \{1, \ldots, M\} \, : \, I_N(P_{X_q Y_q|A_q B_q  w}) >
    I_N^*\} \, .
  \end{align*}
  Let us for the moment assume that, in Step~\ref{step:output}, the
  protocol $\Pi_N$ chooses $f \in \cT$ such that $f \neq \cF$. Then,
  because of assumption~NS, it follows from~\eqref{eq:6}
  that the distance, $\eps'(f)$, of the output bit $R$ from uniform is
  bounded by
  \begin{align*}
     \eps'(f) \leq \frac{1}{2} I_N^* \, .
  \end{align*}
  By the definition of $I_N^*$ this bound tends to zero as $N$
  tends to infinity.  Consequently, for any fixed $\eps'$ and for
  sufficiently large $N$, the output of $\Pi_N$ is $\eps'$-free,
  except if $f \in \cF$.
 
  To conclude the proof of the first claim, we now consider the
  probability, $P(f \in \cF)$, that $f \in \cF$, for $f$ chosen at
  random from the set $\cT$ as in Step~\ref{step:output} (using
  $\eps$-free bits). We will show that whenever $P(f \in \cF)$ is
  non-negligible, then the probability $P(R=\perp)$ that the protocol
  detects a problem in Step~\ref{step:check} and aborts is almost $1$.

  Since $f$ is chosen at random from the set $\cT$, which has size at
  least $M/N$, any possible choice (using $\eps$-free bits) occurs
  with probability at most $(1/2 + \eps)^{\log(M/N)}=(M/N)^{\log(1/2 +
    \eps)}$.  By the union bound, we thus have
  \begin{align} \label{eq_Fsize}
    P(f \in \cF)  \leq |\cF \cap \cT|  (M/N)^{\log(1/2 +  \eps)} \, .
  \end{align} 
  To bound the probability, $P(R=\perp)$, we consider for any
  $q \in \cF\cap \cT$ the probability, $P(\mathrm{abort}|q)$, that the test in
  Step~\ref{step:check} fails for this particular $q \in \cF\cap \cT$, 
  \begin{align*}
    P(\mathrm{abort}|q)
  =
   P_{A_q B_q}(a_0, b_0) P(X_q=Y_q|a_0,b_0) + \!\!\!\!\sum_{\genfrac{}{}{0pt}{}{a,b}{|a-b|=1}}
P_{A_q  B_q}(a,b)  P(X_q\neq Y_q|a,b) \, ,
  \end{align*}
  where $a_0=0$ and $b_0=2N-1$, and $P_{A_q B_q}(a,b)$ is the
  probability that a particular pair $(A_q, B_q) = (a,b)$ is chosen,
  given that $q \in \cT$. (All these probabilities should be
  understood as conditional probabilities, that, as well as being
  conditioned on $w$, are also conditioned on the values $A_{q'}$,
  $B_{q'}$, $X_{q'}$, and $Y_{q'}$, for $q' < q$.) We now use that the
  probability of any particular pair $(a, b)$ being chosen using
  $\eps$-free bits is at least $(1/2-\eps)^{2\log (2N)}$, whereas the
  total probability of choosing any of the $2 N$ possible neighbouring
  pairs $(a, b)$ is upper bounded by $2N (1/2+\eps)^{2\log N}$. We
  thus have
  \begin{align*}
    P_{A_q B_q}(a, b) \geq \frac{(1/2-\eps)^{2\log N}}{2 N
      (1/2+\eps)^{2\log N}} \, .
  \end{align*}
  Combining this with the above expression for $P(\mathrm{abort}|q)$
  leads to
  \begin{align} \label{eq_probdetection}
    P(\mathrm{abort}|q) 
  >
     \frac{(1/2-\eps)^{2\log N}}{2 N (1/2+\eps)^{2\log N}}  I_N(P_{X_q
       Y_q|A_q B_q w})
  \geq 
    \frac{(1/2-\eps)^{2\log N}}{2 N (1/2+\eps)^{2\log N}}  I_N^*  =:
    \bar{p}  \, .
   \end{align} 
   The probability that the protocol does not abort in
   Step~\ref{step:check} is a bound on $P(R\neq\bot)$, and so we have
  \begin{align*}
    P(R\neq\bot) \leq \prod_{q \in \cF \cap \cT}
    \bigl(1-P(\mathrm{abort}|q)\bigr) \leq (1-\bar{p})^{|\cF \cap
      \cT|} = \bigl((1-\bar{p})^{1/\bar{p}}\bigr)^{|\cF\cap \cT|
      \bar{p}} \leq e^{-|\cF \cap \cT| \bar{p}} \, .
  \end{align*}
  Using~\eqref{eq_Fsize} and~\eqref{eq_probdetection} we can bound the
  exponent on the right hand side by
  \begin{align*}
    |\cF \cap \cT| \bar{p} & \geq 
    (M/N)^{-\log(1/2 + \eps)} I_N^* (1/2-\eps)^{2\log N}
    (1/2+\eps)^{-2\log N} (2 N)^{-1} P(f \in \cF) \\
    & \geq
      (M/N)^{-\log(1/2 +  \eps)} N^{2 \log(1/2-\eps)
       - 2 \log(1/2+\eps)}  I_N^*(2N)^{-1} P(f \in \cF)  \\
    & \geq  N^{-399/100 \log(1/2 +  \eps) + 2 \log(1/2-\eps)-101/100} P(f \in \cF)  
\, , 
  \end{align*}
  where we have inserted $I_N^*=2 N^{-1/100}$ and $M=N^{299/100}$. The
  exponential term on the right hand side grows with increasing $N$
  for $\eps<0.058$. This implies that, unless $P(f\in\cF)$ tends to
  zero, the probability of abort, $P(R=\perp)$, tends to $1$ as $N$
  tends to infinity. This concludes the proof of the first claim.

  To establish the second claim, first note that the expected size of
  $|\cT|$ is $2M/N$, and hence, for large $N$, the probability of
  abort in Step~\ref{step:modS} tends to zero.  In addition, quantum
  correlations can achieve a value of $I_N$ that scales like $1/N$,
  and hence the probability of failing the test in
  Step~\ref{step:check} scales like $1/N^2$ for each member of $\cT$.
  Thus, since $|\cT|\leq 3M/N$, with the above choice of $M$, the
  average number of detections in the case with perfect quantum states
  scales like $N^{-1/100}$, which tends to $0$ for large $N$.  Thus,
  the protocol will almost never abort if correctly implemented with
  quantum states.
\end{proof}

\section*{Limitation of using chained Bell correlations}
In the following we show that, using the above approach based on
chained Bell correlations, the threshold on $\eps$ in Theorem~1 cannot
be made arbitrarily small. To do so, we prove that if $\eps$ is above
a certain value, then these correlations admit a classical
explanation.

We first note that a classical strategy can always appear to satisfy
the correlations (lead to a measured value of $I_N=0$) if one pair of
$A$, $B$ values present in the definition of $I_N$ is known not to
occur.  Furthermore, using the best possible classical strategy, for
each $W=w$, either $P(X=Y|a_0,b_0,w)$ or one of $\{P(X\neq
Y|a,b,w)\}_{|a-b|=1}$ will equal~$1$ and all the others will be~$0$.
Therefore, the optimal classical strategy involves a setup in which
the term that equals $1$ corresponds to the pair $(a,b)$ with the
minimum probability of occurring, which can be set using $W$. (In the
following we use $W=(a,b)$ to indicate the pair of $a$ and $b$ values
that are least likely to occur.)  For $N=2^r$, we have
$\min_{a,b}P_{AB|w}(a,b)=(\frac{1}{2}-\eps)^{2r}$ and we assume the
minimal pair is chosen uniformly over the pairs $(a,b)$ in $I_N$ (this
makes it easiest to recreate the correlations using a classical
strategy, i.e., in a cryptographic picture, it gives the greatest
power to the adversary).  We hence have 
\begin{eqnarray*}
  P(X\neq Y|a,b)&=&\sum_w P(X\neq Y,W=w|a,b)\\
  &=&P(W=(a,b)|a,b)\\
  &=&\frac{P(W=(a,b))P(a,b|W=(a,b))}{P(a,b)}\\
  &=&\frac{2^{-(r+1)}(\frac{1}{2}-\eps)^{2r}}{2^{-2r}}=2^{r-1}\left(\frac{1}{2}-\eps\right)^{2r}
\end{eqnarray*}
for $(a,b)\neq(a_0,b_0)$, and the same value is obtained for
$P(X=Y|a_0,b_0)$.  The value of $I_{2^r}$ that
would be observed is then 
$(1-2\eps)^{2r}$.

In order to be consistent with quantum theory, this should be at most
$2^{r+1}\sin^2\frac{\pi}{2^{r+2}}$, i.e.\
$$\frac{1}{2}-\eps\leq\frac{1}{\sqrt{2}}\left(2\sin^2\frac{\pi}{2^{r+2}}\right)^{\frac{1}{2r}}\,
.$$ This function is decreasing in $r$, so, in order to achieve
arbitrarily free output bits with the largest $\eps$, we should use
the largest possible $r$.  For large $r$, the right hand side
approaches $(\pi^2/8)^{1/2r}/(2\sqrt{2})$, hence, in the limit
$r\rightarrow\infty$, consistency with quantum theory is achievable
for $\eps\geq(1-1/\sqrt{2})/2\approx 0.146$.

The above places limitations on when free randomness amplification is
possible using chained Bell correlations (we cannot expect to improve
the quality of bits from sources of $\eps$-free bits with $\eps\geq
0.146$, using these correlations).  Note that related limitations on
using bipartite correlations (in the context of demonstrating
non-locality) have been found in other recent
work~\cite{KPB,Hall_FW,BarrettGisin,Hall_KS}.

\section*{Possible route to proving Conjecture~1}
We hint that one may be able to establish the truth of Conjecture~1
using GHZ states~\cite{GHZ}.  That GHZ correlations may be more useful
for this task comes from the following observations about GHZ
correlations: for any $0\leq\eps<\frac{1}{2}$, $\eps$-free bits are
sufficient to demonstrate non-locality for these correlations (in
contrast to the bipartite case, whose limitations were described
above).  We first outline a few important properties of these
correlations.

For $M$ parties, GHZ correlations are those generated by measuring
each part of the state $\left(\ket{0\ldots 0}-\ket{1\ldots 1}\right)
/\sqrt{2}$ in either the $\{\ket{+}_x,\ket{-}_x\}$ or
$\{\ket{+}_y,\ket{-}_y\}$ basis (where
$\ket{\pm}_x=(\ket{0}\pm\ket{1})/\sqrt{2}$ and
$\ket{\pm}_y=(\ket{0}\pm i\ket{1})/\sqrt{2}$), and label the outcomes
$1$ and $-1$. For convenience, we denote the inputs corresponding to
these bases $0$ and $1$ respectively.  These correlations have the
property that certain output combinations are impossible.  For
example, if $M=3$ and all three parties input $0$ the product of the
outcomes is always $-1$.  We now consider a classical strategy, which
corresponds to an assignment of outputs to each input (this assignment
may depend on some additional variables, $W$).  We label the bits
assigned to the $i$th output by $x_i^0\in\pm 1$ and $x_i^1\in\pm 1$,
where the superscript refers to the possible inputs $A_i=0$ or
$A_i=1$.

In order to mimic the quantum correlations, the classical output bits
need to satisfy $x_1^0x_2^0x_3^0=-1$, $x_1^0x_2^1x_3^1=1$,
$x_1^1x_2^0x_3^1=1$ and $x_1^1x_2^1x_3^0=1$.  It is easy to see that
this is impossible (for example, taking the product of all three
equations yields $(x_1^0x_2^0x_3^0x_1^1x_2^1x_3^1)^2=-1$).  However,
there are classical strategies which satisfy 3 of these relations (for
example, when each output is $1$, independent of its input).

We now imagine choosing measurements to perform on tripartite GHZ
states using bits that are $\eps$-free.  As mentioned above, for any
classical strategy, there is at least one combination of inputs that
yields an incorrect set of outputs.  Using the $\eps$-free source of
randomness in three places, the probability of such an input is
$(\frac{1}{2}-\eps)^3$.  Hence, for any $\eps<\frac{1}{2}$, the
presence of a classical strategy will eventually be noticed as more
tests are performed.  We conclude that non-locality can be verified
with $\eps$-free bits provided $\eps<\frac{1}{2}$ (i.e.\ the bits are
not completely correlated with $W$).

Nevertheless, it does not follow that the outputs of such measurements
are completely free, and, in fact, it is easy to see that they may not
be.  One set of no-signalling correlations that satisfy all the GHZ
relations is realized by having a deterministic output (conditioned on
$W$) for one of the parties, and a non-local box~\cite{Cirelson93,PR}
shared between the remaining two~\cite{BM}.  Using these correlations,
there is always one output that is deterministic and hence not free.

However, we suggest that arbitrarily free bits may be generated from
partially free ones using an $M$-party GHZ state for large $M$.  The
partially free source of randomness is used in $M$ places to choose
measurements on each part of the state in either of the two bases
specified above.  Then, if the outputs satisfy the $M$-party GHZ
relations, one of the randomness sources is used to pick one of the
$M$ output bits at random.  The idea is that, in the limit of large
$M$, this output is arbitrarily close to being perfectly free, except
with very small probability.  However, it may turn out that other
states and measurements are required in order to prove
Conjecture~1.

A corollary of the above is that if the measurement devices are
restricted to be quantum (rather than arbitrary no-signalling, i.e.\
we trust that the measurement devices are limited by quantum theory,
but not what they are doing internally), $\eps$-free bits for any
$0\leq\eps<\frac{1}{2}$ can be used to generate arbitrarily free bits.
This follows from the observation that the only quantum states that
perfectly obey the tripartite GHZ relations are (up to local unitary
operations) GHZ states~\cite{ColbeckThesis,CK2}, from which perfect
randomness can be derived by taking any of the three outputs. Hence, a
set of quantum measurement devices that never deviate from the GHZ
relations (using measurements chosen with an $\eps$-free source) also
generate perfectly free randomness.


\end{widetext}

\end{document}